\documentclass[prd, preprint, amsfonts,amssymb]{revtex4-1}
 \usepackage{hyperref}
 \usepackage{graphicx}
 \usepackage{amsmath}
 \usepackage{amssymb}
  \usepackage{enumerate}
  \usepackage{bbm}
  \usepackage{pbox}
  \usepackage{csquotes}
\usepackage{epsf,epsfig,amsmath,amssymb,dsfont}
\usepackage{amsthm,mathrsfs} %mathematical symbols
\usepackage{graphicx} %includes images
\usepackage{float}
\usepackage{slashed}
\usepackage{float}
\usepackage{hyperref} %per avere i link nel testo
\usepackage{multirow}
\usepackage{mathtools}
\usepackage{color}
\usepackage{cancel}
  \usepackage{bbold}
\newcommand{\be}{\begin{eqnarray}}
\newcommand{\ee}{\end{eqnarray}}
 \usepackage[framemethod=TikZ]{mdframed}
 
\begin{document}
  \title{How black holes store  information in high-order correlations }
  \author {Charis Anastopoulos}
  \email{anastop@upatras.gr}
  \author{Konstantina Savvidou}
  \email{ksavvidou@upatras.gr}
  \affiliation{ Department of Physics, University of Patras, 26500 Greece}
\date{\today }

\begin{abstract}

 We explain  how Hawking radiation  stores significant amount of information in high-order correlations of quantum fields. This information can be retrieved by  multi-time measurements on the quantum fields close to the black hole horizon. This result requires no assumptions about quantum gravity, it takes into account the differences between Gibbs's and Boltzmann's accounts of thermodynamics, and it clarifies misconceptions about key aspects of Hawking radiation and about informational notions in QFT.
 \\
\\
  \textit{ Honorable mention at the Gravity Research Foundation 2020 Awards for Essays on Gravitation.}
\end{abstract}
\maketitle

 Stephen Hawking’s discovery  that black holes radiate thermally implies that black holes eventually evaporate. The semiclassical analysis of the evaporation process leads to the conclusion that the final state should be maximally mixed. This suggests  an evolution law that takes pure states to  mixed states, i.e., non-unitary time evolution.  Most quantum gravity research programs presuppose unitary time evolution, hence,  many researchers   refer to the prediction above as the {\em paradox} of black-hole information loss.

 In this essay, we will explain  how Hawking radiation can store significant amount of information in higher-order correlations of the quantum field. This result follows from Quantum Field Theory (QFT) in curved spacetime, without any assumptions about quantum gravity. It is fully compatible with the semiclassical analysis of black hole evaporation, and it takes into account the differences between Gibbs's and Boltzmann's accounts of thermodynamics.  It requires the resolution of common misconceptions  about key results on Hawking radiation and  about informational notions in QFT. Non-unitarity arises from the misguided attempt to define evolving single-time quantum states in a non-globally hyperbolic spacetime, i.e., in a system that lacks a global notion of time. But it does not necessitate quantum information loss.

A crucial fact in the analysis of black hole information loss is that the reduced quantum state of the field far from the horizon  is asymptotically Gibbsian at the Hawking temperature $T_H$. This is the content of the Hawking-Wald (HW) theorem \cite{Hawk1, Wald1}. It implies that the values of all physical observables outside the black hole are distributed thermally, at late times. There are no correlations  between quanta of different field modes. This conclusion is important because it implies that when the black hole totally evaporates, it leaves only Hawking radiation behind, which is described by a Gibbsian quantum state.  This is a maximally mixed state, hence, the process of black hole formation and evaporation ostensibly involves tremendous information loss \cite{Hawk76}.

 For macroscopic black holes, the rate of Hawking radiation is very small, so we expect that  the black hole geometry changes in a quasi-stationary way. Therefore, the semiclassical approximation will be good until the black hole shrinks near the Planck mass. Quantum gravity effects at this late stage cannot affect the radiation that has already been emitted. It follows that the final state cannot be very different from what the HW theorem predicts, i.e., a  maximally mixed state with no correlations and no capacity to carry information.  Hence,
the conclusion of information loss is not affected by the consideration of backreaction.

 \begin{figure}
    \centering

 \includegraphics[width=0.65\textwidth]{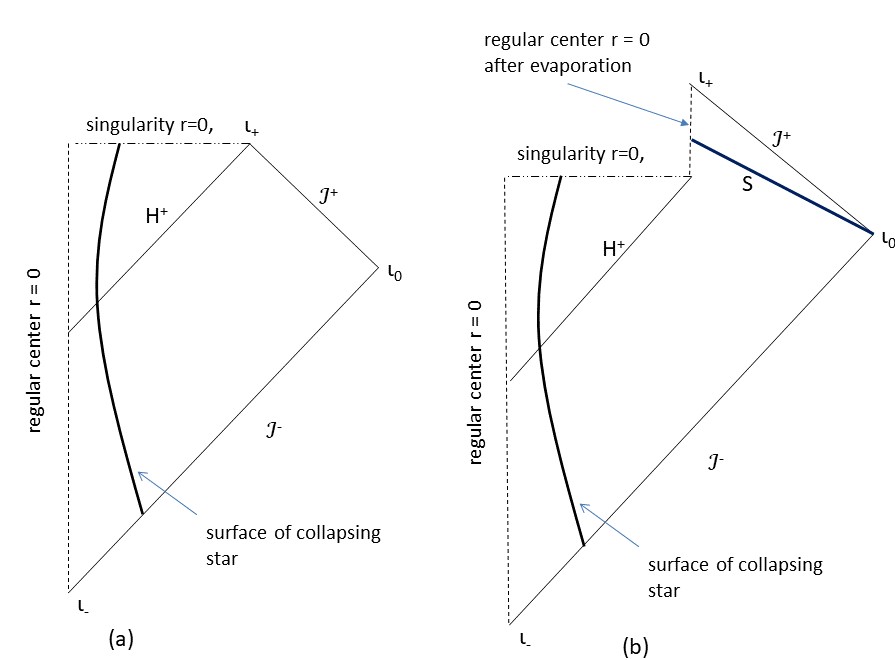}

    \caption{ Penrose diagrams for (a) a  black hole formed from gravitational collapse and (b)  a black hole that is formed from collapse and then evaporates. }
\end{figure}

Note that the above analysis presupposes a Gibbsian rather than a Boltzmannian description of  Hawking radiation thermodynamics. In Gibbs's description, equilibrium is defined at the level of microstates.  In Boltzmann's approach, equilibrium refers only to the behavior of macrostates---see, Table 1. The crucial difference is that Boltzmann's theory does not constrain observables that are not distinguishable at the macrostate level.

\begin{table}
 \begin{tabular}{|c|c|c|}
   \hline
   \small
  {} & {\bf Gibbs} & {\bf Boltzmann} \\
   \hline \hline
 {\bf definition} & statistical ensemble & individual systems \\
 \hline
 {\bf description} & microstates & macrostates \\
 \hline
 {\bf gas of $N$ particles} &\pbox{20cm}{probability distribution \\ (canonical) on state
   \\ space $\Gamma = {\pmb R}^{6N}$}& \pbox{20cm}{probability distribution \\ (Maxwell-Boltzmann) \\ on single-particle  state \\ space  $\mu = {\pmb R}^{6}$} \\
    \hline
   {\bf QFT} &\pbox{20cm}{fixes complete  hierarchy  \\ of  correlation  functions  \\} &\pbox{20cm}{fixes  two-point  functions} \\
   \hline
 \end{tabular}
 \caption{Gibbsian vs. Boltzmannian approach to statistical mechanics.}
\end{table}

 The HW theorem is much less powerful than what has been commonly assumed, in the sense that it precludes only one type of correlations, namely, correlations that refer to a single moment of time. It makes no statement about multi-time correlations, i.e., correlations associated to measurements at different moments of time. In Ref. \cite{AnSav20}, we showed that, at the level of QFT in a background black hole spacetime,  multi-time correlations in Hawking radiation  are not thermal and  they preserve significant memory from the history of Hawking quanta. We conclude therefore that  they
 can carry non-trivial amounts of information.

 The HW theorem treats the quantum field as a bipartite system: one part consists of field states at the horizon ${\cal H}^+$, and one part consists of states at the future null infinity $\mathcal {I}^+$---see, Fig. 1.a. When we trace out the states at ${\cal H}^+$, we are left with the {\em reduced density matrix} for the states at $\mathcal {I}^+$. The latter density matrix is mixed---in fact, thermal---because of strong entanglement between the two subsystems.

In general, a time-evolving reduced density matrix misrepresents the probabilities for multi-time measurements in the associated subsystem.
This is a well-known fact from the theory of open quantum systems that has not been taken into account in past discussions of information loss. A simple extension of the HW theorem suffices to show that multi-time correlations  cannot be expressed solely in terms of the degrees of freedom at $\mathcal {I}^+$.  They also involve   field states from ${\cal H}^+$, hence, they cannot be thermal \cite{AnSav20}.

%This result greatly affects our understanding of information balance in the processes of black hole formation and evaporation.

Past analyses of   information loss ignored   multi-time QFT measurements, as they focused mainly on the S-matrix description of the field. By contrast, multi-time  field measurements
 are ubiquitous in quantum optics. They are described by Glauber's  photo-detection theory \cite{Glauber} that defines the higher-order coherences of the quantum electromagnetic field. They account for  phenomena like the Hanbury-Brown-Twiss effect,   photon bunching and anti-bunching.

In recent years, we developed a new formalism for QFT measurements  \cite{QTP}   that greatly generalizes Glauber's theory. We call this method the Quantum Temporal Probabilities (QTP) method, as its original motivation was to provide a general framework for  {\em temporally extended} quantum observables.
In QTP, the probability density $W_n(X_1, X_2, \ldots, X_n)$ for $n$ particle detection events at spacetime points $X_i$  is a linear functional of the field $2n$-point function
\be
G^{(2n)}(X_1, X_2, \ldots, X_n; X_1', X_2', \ldots, X_n') :=  Tr \left\{{\cal T}\left[ \hat{O}(X_n) \ldots \hat{O}(X_2)  \hat{O}(X_1)   \right] \hat{\rho}_0  \right.
\nonumber\\ \left.
\times  \bar{\cal T}\left[ \hat{O}(X_1') \hat{O}(X_2') \ldots  \hat{O}(X_n')  \right]  \right\},  \label{nmpt}
\ee
where ${\cal T}$ stands for time-ordering,  $\bar{\cal T}$ for reverse-time-ordering, and $\hat{\rho}_0$  is the field initial state. The local composite operator $\hat{O}(X)$ defines the channel of interaction between the quantum field and the detector.
These correlation functions are not used in S-matrix theory but they appear in  the Schwinger-Keldysh  formulation of QFT  \cite{CTP}.

Full information about a quantum field is contained in the hierarchy of $N$-point correlation functions, for all $N = 1, 2, \ldots, \infty$.   The main achievement of QTP is that it defines a hierarchy of probability functions for multi-time measurements from the
 QFT  correlation hierarchy \cite{QTP}---see, Fig. 2.

 \begin{figure}
    \centering

 \includegraphics[width=1.05\textwidth]{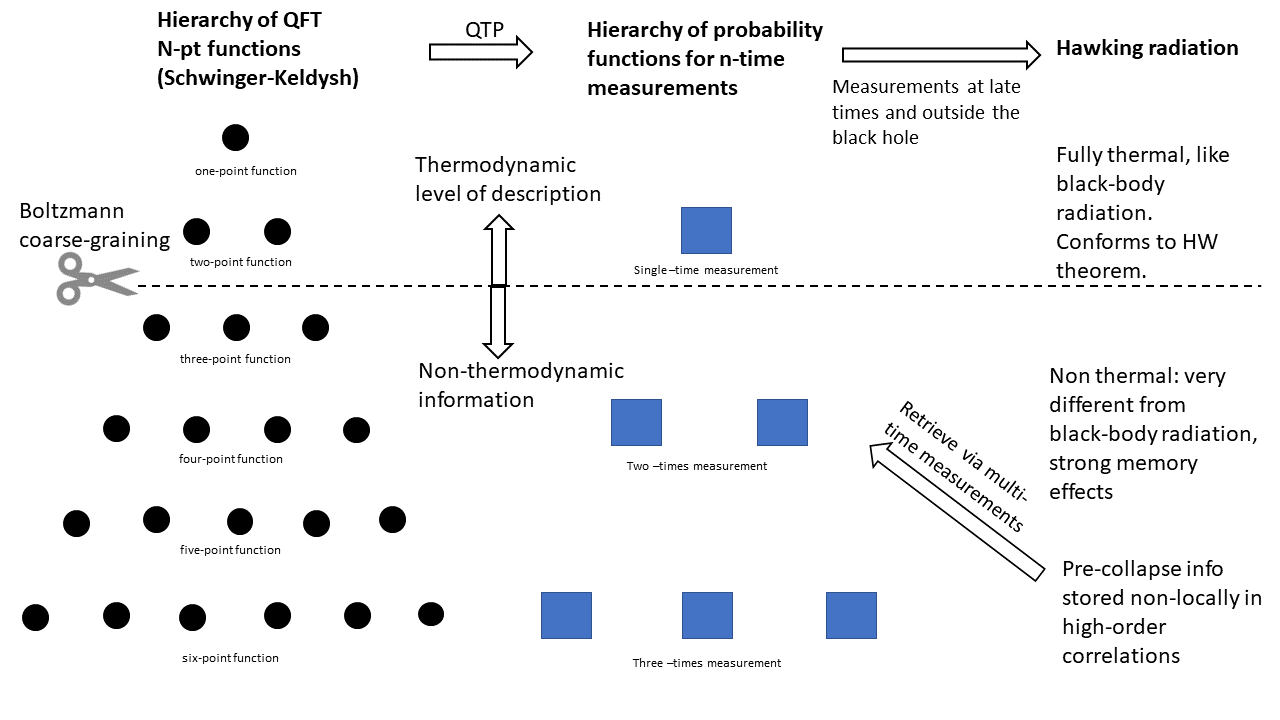}

    \caption{QFT correlation hierarchy, Boltzmann-coarse graining and information. }
\end{figure}

 A  field  description  that considers anything less than the full hierarchy is  {\em coarse-grained}. Coarse-graining is usually associated with loss of information.  It is essential for deriving  the second law of thermodynamics from microscopic theories. In particular, Boltzmann's equation in QFT follows from a truncation of the correlation hierarchy at the level of the two-point functions \cite{CTP}. The truncated hierarchy defines the {\em thermodynamic level of description}, i.e., Boltzmann's macrostates. The second law of thermodynamics is essentially the statement that information is typically lost from this level, as it is transferred to higher-order correlation functions.

 The analysis of Hawking radiation through multi-time measurements outside the black hole revealed the following pattern: (i)
  Single-time measurements of the field outside the black hole `capture' information from the two-point correlation functions. The associated probabilities are thermal, in accordance with the HW theorem.
(ii) Multi-time measurements of the field  outside the black hole  `capture' information from the $N > 2$ correlation functions.  The associated probabilities  do not behave thermally, and they support significant correlations.

 Therefore,  Hawking radiation behaves thermally only with respect to Boltzmann's  coarse-graining. This strongly contrasts the usual understanding of Hawking-radiation thermodynamics in terms of a
 Gibbsian quantum state. In the QFT context,  a Gibbs state  specifies the full hierarchy of correlation functions, while Boltzmann's theory specifies only two-point correlation functions.  It is widely agreed that Boltzmann's description is more fundamental, because it applies to individual closed systems and not statistical ensembles. Furthermore,  Gibbs's theory  does not naturally account for non-equilibrium thermodynamics and   the second law \cite{BoltzGib}.

Hence,   the loss of information in black hole evaporation  is identical to that in non-equilibrium statistical mechanics, in the sense that  information is transferred to thermodynamically inaccessible {\em non-local} degrees of freedom. This information can only be retrieved  by multi-time measurements.

Our results are consistent with the  semi-classical analysis of black hole evaporation. The quantum backreaction on  the spacetime metric is driven by the expectation value $\langle \hat{T}_{\mu \nu}\rangle$ of
the quantum stress energy tensor $\hat{T}_{\mu \nu}$, defined in terms of the field two-point function. Backreaction transfers information to inaccessible degrees of freedom through other channels---including higher-order correlations of  $\hat{T}_{\mu \nu}$. The key point here is that this information transfer takes place independently of the strength of $\langle \hat{T}_{\mu \nu}\rangle$, i.e., even if backreaction has little effect on spacetime geometry.

  %The transfer of information to inaccessible degrees of freedom is a continuous and persistent  process that does not require significant transfer of energy.  In the black hole context, the change in geometry through backreaction is driven by the quantum stress energy tensor. The latter  is expressed in terms of the field two-point function, i.e., it is a variable at the thermodynamic level of description. Hence, backreaction can
%dramatically change the information balance in the higher order correlation functions, while preserving the results of the semi-classical analysis until the later stages of black hole evaporation.

We proved  that   pre-collapse information is not necessarily lost. We believe that it is stored into higher-order correlations through backreaction, and it is accessible by multi-time measurements close to the horizon. For example, correlations at $ 2M < r < 3M$, in a Schwarzschild spacetime of mass $M$,  carry the strongest memories  about the past history of Hawking quanta \cite{AnSav20}.  Multi-time measurements after the collapse  allow us to {\em retrodict}  pre-collapse properties of the system. In this sense, black holes have quantum informational hair.

Finding exactly how much information is stored in Hawking  radiation correlations requires a first-principles extension of  quantum information theory to relativistic QFTs that is currently unavailable. However,  even  with maximal retrieval of pre-collapse information,   unitarity will not be restored. Non-unitarity originates from the fact that the notion of an instant of `time'  does not exist after evaporation, i.e., the spacelike surface $S$ of Fig. 1.b is not  a Cauchy surface. What breaks down is the notion of the single-time quantum state;  it is a conceptual mistake to focus on unitarity.    We believe that generalizations of quantum theory that  are based on the notion of history \cite{histories}--- treating single-time quantum states as derived concepts---are more appropriate for the physics of black hole evaporation \cite{hartle98} and for quantum gravity \cite{Sav03, Sav10}.

  \end{document}